\documentclass[aps,prd,nofootinbib,amsmath,amssymb,showpacs,showkeys,superscriptaddress,twocolumn,10pt]{revtex4}
\usepackage{amsmath}
\usepackage{amsfonts}
\usepackage{txfonts}
\usepackage{graphicx}
\usepackage{dcolumn}
\usepackage{bbm}
\usepackage{amssymb}
\usepackage{latexsym}
\usepackage{CJK}
\usepackage{titlesec}
\usepackage{color}
\usepackage[colorlinks=true, linkcolor=red, citecolor=blue]{hyperref}

\begin{document}

 \bibliographystyle{unsrt}

\title{Strong gravitational lensing constraints on holographic dark energy}

\author{Jing-Lei Cui}
\affiliation{Department of Physics, College of Sciences, Northeastern University,
Shenyang 110819, China}
\author{Yue-Yao Xu}
\affiliation{Department of Physics, College of Sciences, Northeastern University,
Shenyang 110819, China}
\author{Jing-Fei Zhang}
\affiliation{Department of Physics, College of Sciences, Northeastern University,
Shenyang 110819, China}
\author{Xin Zhang\footnote{Corresponding author}}
\email{zhangxin@mail.neu.edu.cn} \affiliation{Department of Physics, College of Sciences,
Northeastern University, Shenyang 110819, China}
\affiliation{Center for High Energy Physics, Peking University, Beijing 100080, China}

\begin{abstract}
Strong gravitational lensing (SGL) has provided an important tool for probing galaxies and cosmology. In this paper, we use the SGL data to constrain the holographic dark energy model, as well as models that have the same parameter number, such as the $w$CDM and Ricci dark energy models. We find that only using SGL is difficult to effectively constrain the model parameters. However, when the SGL data are combined with CBS (CMB+BAO+SN) data, the reasonable estimations can be given and the constraint precision is improved to a certain extent, relative to the case of CBS only. Therefore, SGL is an useful way to tighten constraints on model parameters.
\end{abstract}



\maketitle

\renewcommand{\thesection}{\arabic{section}}
\renewcommand{\thesubsection}{\arabic{subsection}}
\titleformat*{\section}{\flushleft\bf}
\titleformat*{\subsection}{\flushleft\bf}
\section {Introduction}
Dark energy (DE) has been proposed to explain the accelerating expansion of the universe found by various observations~\cite{Riess:1998cb,Perlmutter:1998np,Tegmark:2003ud,Eisenstein:2005su,
Spergel:2003cb}.
We actually know little about its nature other than the fact that it is gravitationally repulsive and contributes about $70\%$ of the cosmic energy density. Many cosmologists suspect that the identity of DE is the cosmological constant, and the corresponding DE model is dubbed as $\Lambda$CDM, which fits the observational data quite well. However, the $\Lambda$CDM model confronts with the severe theoretical problems, i.e.,``fine-tuning" and ``cosmic coincidence" puzzles~\cite{Sahni:1999gb,Padmanabhan:2002ji,Bean:2005ru,Copeland:2006wr,Sahni:2006rde,limiao}.
Hence, cosmologists constructed a host of theoretical/phenomenological DE models~\cite{quintessence,Ratra:1987rm,Caldwell:1999ew,Zhang:2005kj,Zhang:2005rg,Sen:2002in,ArkaniHamed:2003uy,
ghost,Wei:2007ty,Kamenshchik:2001cp,Bento:2002ps,Zhang:2004gc,Chevallier:2000qy,Linder:2002et,Ma:2011nc,Li:2012via,Li:2012dt} in terms of the physics different from $\Lambda$CDM.
For the recent progress in DE field, see, e.g.,~\cite{Li:2014eha,Li:2014cee,sci01,zjf2014epjc,sci02,zjf2015jcap,sci03,Geng:2014hoa,Zhang:2013lea,Zhang:2014ifa,sci04}.

The holographic dark energy (HDE) model~\cite{Li:2004rb} is a dynamical DE model based on the holographic principle. Cohen et al.~\cite{Cohen:1998zx} pointed out that the total energy of a system with size $L$ should not exceed the mass of a black hole with the same size, i.e., $L^3\rho _{\rm{de}}\lesssim LM^2_{\rm{P}}$. This energy bound leads to the density of holographic dark energy,
\begin{equation}
\rho _{\rm{de}} = 3c^2M^2_{\rm{P}}L^{-2},\label{eq1}
\end {equation}
where $c$ is a dimensionless parameter characterizing some uncertainties in the effective quantum field theory, $M_{\rm{P}}$ is the reduced Planck mass defined by $M^2_{\rm{P}} = (8\pi G)^{-1}$, and $L$ is the infrared (IR) cutoff in the theory. Li~\cite{Li:2004rb} suggested that the IR cutoff $L$ should be given by the event horizon of the universe, defined as
\begin{equation}
L=a(t)\int_t^\infty\frac{dt'}{a(t')}=a\int_a^\infty\frac{da'}{Ha'^2},\label{eq2}
\end {equation}
where $a$ is the scale factor of our universe, the Hubble parameter $H=\dot{a}/a$, and the dot denotes the derivative with respect to time $t$. This yields the HDE model studied in this paper, which is a very competitive candidate for DE, attracting lots of studies~\cite{Huang:2004mx,Zhang:2005yz,hologhost,Zhang:2006av,holotach,Zhang:2007an,Ma:2007av,Li:2008zq,Cui:2010dr,Zhang:2009xj,Li:2010cj}.

 To constrain the parameter space of various DE models, the observational data including cosmic microwave background (CMB), baryon acoustic oscillations (BAO), type Ia supernova (SN) and so on, have been used~\cite{Zhang:2005hs,Zhang:2007sh,Chang:2005ph,Huang:2004wt,Wang:2004nqa,
Ma:2007pd,Li:2009bn,Li:2009zs,Li:2009jx,Li:2013dha,Wang:2012uf,Duan:2011jj}. However, to obtain tighter constraints on parameters, we need more cosmological probes, for example, the strong gravitational lensing (SGL).
SGL can provide the information about the proper angular diameter distances between lens and source and between observer and source. Since these distances depend on cosmological geometry, we can use their ratio to constrain the parameters in cosmological model. Furthermore, the strong lens systems produced by massive galaxies are fairly useful to constrain the statistical properties of galaxies.
 After the pioneer SGL in Q0957+561~\cite{Walsh:1979nx} was discovered, SGL has increased to enough number and developed into an important astrophysical tool for probing both galaxies (their structure, formation and evolution)~\cite{Zhu:1997jf,Mao:1997ek,Jin:1999we,Keeton:2001gb,Kochanek:2001sj,Ofek:2003sp,Treu:2005aw} and cosmology~\cite{Zhu:2000ee,Zhu:2000wx,Chae:2002uf,Chae:2004jp,Mitchell:2004gw,Zhu:2008sg,Zhu:2007tm,Cao:2012ja}.

In this paper, we constrain the HDE model with 73 gravitational lens data points from Sloan Lens ACS (SLACS) and Lens Structure and Dynamics (LSD) survey released in Ref.~\cite{Cao:2011aq}, which were also used in the constraints on the $f(R)$~\cite{Liao:2012ws}, $f(T)$~\cite{Wu:2014wra} and $\phi$CDM~\cite{Chen:2013vea} models. Besides, the combination of SGL with other geometrical measurements including CMB, BAO and SN (together as ``CBS") is also considered. Furthermore, for comparison, we also constrain other DE models with the same parameter number, such as $w$CDM (with the constant equation of state $w$) and Ricci dark energy (RDE)~\cite{Gao:2007ep} models, by using the same data.

We arrange the paper as follows. In Sec.~\ref{sec.2}, we briefly introduce the data and analysis methods. We present the constraint results of the HDE model in Secs.~\ref{sec.3}. The constraint results of the other two models are presented in Sec.~\ref{sec.4}. The conclusion is shown in Sec.~\ref{sec.5}.

\section {Data and analysis methods}\label{sec.2}

In this section, we describe the observational data used in this paper, including SGL, CMB, BAO and SN. We first give the definitions of the proper angular diameter distance $D_{\rm{A}}(z)$ and luminosity distance $D_{\rm{L}}(z)$, due to their significance in the geometrical probes. In a spatially flat Friedmann-Robertson-Walker (FRW) universe, we have
\begin{equation}
D_{\rm{A}}(z)=\frac{1}{H_0(1+z)}\int_0^z \frac{dz'}{E(z')},\label{DA}
\end{equation}
\begin{equation}
D_{\rm{L}}(z)=\frac{(1+z)}{H_0}\int_0^z \frac{dz'}{E(z')},\label{DL}
\end{equation}
where $E(z)\equiv H/H_0$ is the dimensionless Hubble expansion rate.

\subsection*{2.1 Strong gravitational lensing}

Gravitational lensing effect refers to that light rays from a distant quasar or galaxy to us will be bent when they travel through gravitational field given by galaxy or galaxy cluster acting as lens. Strong lensing is to produce multiple images of the source, arcs or even Einstein rings. In the lensing analyses, the Einstein radius ($\theta_{\rm{E}}$) is a basic quantity. The formula of $\theta_{\rm{E}}$ in a lensing system modeled by the singular isothermal sphere (SIS) assumption can be expressed as
\begin{equation}
\theta_{\rm{E}} = 4\pi \sigma^2_{\rm{SIS}}\frac{D_{\rm{A}}(z_{\rm{l}},z_{\rm{s}})}{D_{\rm{A}}(0,z_{\rm{s}})},
\end{equation}
where $D_{\rm{A}}(z_{\rm{l}},z_{\rm{s}})$ and $D_{\rm{A}}(0,z_{\rm{s}})$ are the proper angular diameter distances between lens and source and between observer and source respectively, which can be calculated according to Eq.~(\ref{DA}), and $\sigma_{\rm{SIS}}$ is the velocity dispersion of the SIS model. Obviously, according to the distance ratio $D_{\rm{A}}(z_{\rm{l}},z_{\rm{s}})/D_{\rm{A}}(0,z_{\rm{s}})$, the Einstein radius is dependent on the cosmological model and independent on the Hubble constant value. Thus, we can estimate the parameters of the cosmological model through the distance ratio
\begin{equation}
\mathcal{D}(z_{\rm{l}},z_{\rm{s}}) = \frac{D_{\rm{A}}(z_{\rm{l}},z_{\rm{s}})}{D_{\rm{A}}(0,z_{\rm{s}})}=\frac{\int_{z_{\rm{l}}}^{z_{\rm{s}}} dz'/E(z')}{\int_0^{z_{\rm{s}}} dz'/E(z')}.
\end{equation}
In Table 1 of Ref.~\cite{Cao:2011aq}, $80$ data points are listed, including the redshifts of lens and source, $z_{\rm{l}}$ and $z_{\rm{s}}$, as well as the observational $\mathcal{D}^{\rm{obs}}$ and its corresponding uncertainty $\sigma_{\rm{\mathcal{D}}}$. In these data, $70$ strong lensing systems come from SLACS and LSD survey, and $10$ giant arcs samples come from galaxy clusters in the redshift region from $0.1$ to $0.6$. We select $73$ observational data with the selection criterion $\mathcal{D}^{\rm{obs}}<1$ from Table 1 of Ref.~\cite{Cao:2011aq}, which means that the distance between the lens and source should be always smaller than that between the source and observer. Then we use these data to constrain the parameters of DE models by minimizing the following $\chi^2$ function,
\begin{equation}
\chi^2_{\rm{SGL}}=\sum\limits_{i}\frac{(\mathcal{D}^{\rm{th}}_i-\mathcal{D}^{\rm{obs}}_i)^2}{\sigma^2_{\mathcal{D},i}}.
\end{equation}

\subsection*{2.2 Cosmic microwave background}

For CMB data, we use the distance prior data extracted from Planck first data release~\cite{Wang:2013mha}. The distance priors include the shift parameter $R$, the acoustic scale $\ell_{\rm{A}}$ and $\omega_{\rm{b}}\equiv\Omega_{\rm{b0}}h^2$. $R$ and $\ell_{\rm{A}}$ are, respectively, defined as
\begin{equation}
R\equiv \sqrt{\Omega_{\rm{m0}}H^2_0}(1+z_*)D_{\rm{A}}(z_*)
\end{equation}
and
\begin{equation}
\ell_{\rm{A}} \equiv (1+z_*)\frac{\pi D_{\rm{A}}(z_*)}{r_{\rm{s}}(z_*)},
\end{equation}
where $\Omega_{\rm{m0}}$ is the present-day fractional energy density of matter, $D_{\rm{A}}(z_*)$ is the proper angular diameter distance at the redshift of the decoupling epoch of photon $z_*$, and $r_{\rm{s}}(z_*)$ is the comoving sound horizon at $z_*$. Here, $r_{\rm{s}}(z)$ is given by
\begin{equation}
r_{\rm{s}}(z)=H^{-1}_0\int^a_0 \frac{da'}{a'^2E(a')\sqrt{3(1+\overline{R_{\rm{b}}}a')}},\label{rs}
\end{equation}
where $\overline{R_{\rm{b}}}a=3\rho_{\rm{b}}/(4\rho_{\rm{\gamma}})$ with $\rho_{\rm{b}}$ and $\rho_{\rm{\gamma}}$ being the baryon and photon energy densities, respectively. Thus we have $\overline{R_{\rm{b}}}=31500\Omega_{\rm{b0}}h^2(T_{\rm{cmb}}/2.7\rm{K})^{-4}$, where $\Omega_{\rm{b0}}$ is the present-day fractional energy density of baryon, $h$ is the reduced Hubble constant defined by $H=100h$ km/s/Mpc, and $T_{\rm{cmb}}=2.7255\rm{K}$. $z_*$ is given by the fitting formula~\cite{Hu:1995en},
\begin{equation}
z_*=1048[1+0.00124(\Omega_{\rm{b0}}h^2)^{-0.738}][1+g_1(\Omega_{\rm{m0}}h^2)^{g_2}],
\end{equation}
where
\begin{equation}
 g_1=\frac{0.0783(\Omega_{\rm{b0}}h^2)^{-0.238}}{1+39.5(\Omega_{\rm{b0}}h^2)^{-0.763}},~~g_2=\frac{0.560}{1+21.1(\Omega_{\rm{b0}}h^2)^{1.81}}.
\end{equation}

We use the mean values and covariance matrix of $\{\ell_{\rm{A}},R,\omega_{\rm{b}}\}$ in~\cite{Wang:2013mha} obtained from the Planck+lensing+WP data. The $\chi^2$ function for CMB is
\begin{equation}
\chi^2_{\rm{CMB}}=\Delta p_i[{\rm{Cov}}^{-1}_{\rm{CMB}}(p_i,p_j)]\Delta p_j,~~~\Delta p_i=p^{\rm{th}}_i-p^{\rm{obs}}_i,
\end{equation}
where $p_1=\ell_{\rm{A}}$, $p_2=R$ and $p_3=\omega_{\rm{b}}$.

\subsection*{2.3 Baryon acoustic oscillations}

In BAO measurements, the ratio of the effective distance measures $D_{\rm{V}}(z)$ and $r_{\rm{s}}(z_{\rm{d}})$ can be provided. The spherical average gives us the expression of $D_{\rm{V}}(z)$,
\begin{equation}
D_{\rm{V}}(z)\equiv \left [(1+z)^2D^2_{\rm{A}}(z)\frac{z}{H(z)}\right ]^{1/3}.
\end{equation}
$r_{\rm{s}}(z_{\rm{d}})$ can be calculated by using Eq.~(\ref{rs}), and $z_{\rm{d}}$ denotes the redshift of the drag epoch, whose fitting formula is given by~\cite{Eisenstein:1997ik},
\begin{equation}
z_{\rm{d}}=\frac{1291(\Omega_{\rm{m0}}h^2)^{0.251}}{1+0.659(\Omega_{\rm{m0}}h^2)^{0.828}}[1+b_1(\Omega_{\rm{b0}}h^2)^{b_2}],
\end{equation}
where
\begin{equation}
\begin{gathered}
b_1=0.313(\Omega_{\rm{m0}}h^2)^{-0.419}[1+0.607(\Omega_{\rm{m0}}h^2)^{0.674}],\\
b_2=0.238(\Omega_{\rm{m0}}h^2)^{0.223}.
\end{gathered}
\end{equation}

We use $7$ BAO data points from the 6dF Galaxy Survey~\cite{Beutler:2011hx}, the SDSS-DR7~\cite{Beutler:2011hx}, the BOSS-DR11~\cite{Padmanabhan:2012hf} and the ``improved" WiggleZ Dark Energy Survey~\cite{Kazin:2014qga}. The $\chi^2$ function for BAO is
\begin{equation}
\chi^2_{\rm{BAO}}=\Delta p_i[{\rm{Cov}}^{-1}_{\rm{BAO}}(p_i,p_j)]\Delta p_j,~~~\Delta p_i=p^{\rm{th}}_i-p^{\rm{obs}}_i.
\end{equation}
Note that three WiggleZ data are correlated with each other, and the inverse covariance matrix for them is given in~\cite{Kazin:2014qga}.

\subsection*{2.4 Type Ia supernova}

\begin{table}
\caption{Constraint results and corresponding $\chi^2_{\rm{min}}$ for the HDE model by using the SGL and CBS (CMB+BAO+SN) data, as well as the combination of them. In the case of fixed $\Omega_{\rm{m0}}$, we choose $\Omega_{\rm{m0}}=0.27$.}
\label{table1}
\renewcommand{\arraystretch}{1.5}
\centering
\begin{tabular}{cccccccccc}
\\
\hline\hline
Parameter               & $\Omega_{\rm{m0}}$ & $c$ & $\chi^2_{\rm{min}}$\\ \hline

SGL ($\Omega_{\rm{m0}}$ free)              & $0.0244^{+0.1123}_{-0.0136}$
                   & $1.9730^{+0.0270}_{-0.8993}$
                   & $84.1185$&                   \\
SGL ($\Omega_{\rm{m0}}$ fixed)         & $0.27$
                   & $0.8335^{+0.8031}_{-0.3495}$
                   & $87.2689$&                   \\
CBS      & $0.2798^{+0.0083}_{-0.0114}$
                   & $0.6458^{+0.0472}_{-0.0483}$
                   & $430.0321$&                   \\
SGL+CBS  & $0.2783^{+0.0092}_{-0.0100}$
                   & $0.6429^{+0.0515}_{-0.0436}$
                   & $517.6995$&                   \\ \hline
\end{tabular}
\end{table}

We choose to use the first three-year dataset of the Supernova Legacy Survey (SNLS3) consisting of $472$ SN~\cite{Conley:2011ku}. SN Ia data give measurements of the luminosity distance $D_{\rm{L}}(z)$ through that of the distance modulus of each SN. The predicted magnitude of a SN is
\begin{equation}
m_{\rm{mod}}=5\log_{10} [H_0 D_{\rm{L}}(z)]-\alpha (s-1)+\beta \mathcal{C}+\mathcal{M},
\end{equation}
where $D_{\rm{L}}(z)$ can be calculated by Eq. (\ref{DL}), $s$ is the stretch measure of the SN light curve shape, $\mathcal{C}$ is the color measure for the SN, $\alpha$ and $\beta$ are nuisance parameters characterizing the stretch-luminosity and color-luminosity relationships, $\mathcal{M}$ is another nuisance parameter representing some combination of the absolute magnitude of a fiducial SN and the Hubble constant.
$\alpha$ and $\beta$ are free parameters during the cosmology-fitting process. Although some works~\cite{Wang:2013yja,Wang:2013tic,Wang:2014oga,Wang:2013zca} considered $\beta$ changing over time, we treat $\alpha$ and $\beta$ as constants in this work.
Introducing a vector with $N$ components
$\Delta \textbf{m}=\Delta \textbf{m}_B-\Delta \textbf{m}_{\rm{mod}}$ ($ m_B$ is the rest-frame peak $B$ band magnitude of a SN), we have
\begin{equation}
\chi^2_{\rm{SN}}=\Delta \textbf{m}^T\cdot {\textbf{C}}^{-1}\cdot \Delta \textbf{m},
\end{equation}
where $\textbf{C}$ is the $N\times N$ covariance matrix of the SN given by~\cite{Conley:2011ku}.

\begin{figure*}[htbp]
\centering
\includegraphics[scale=0.5]{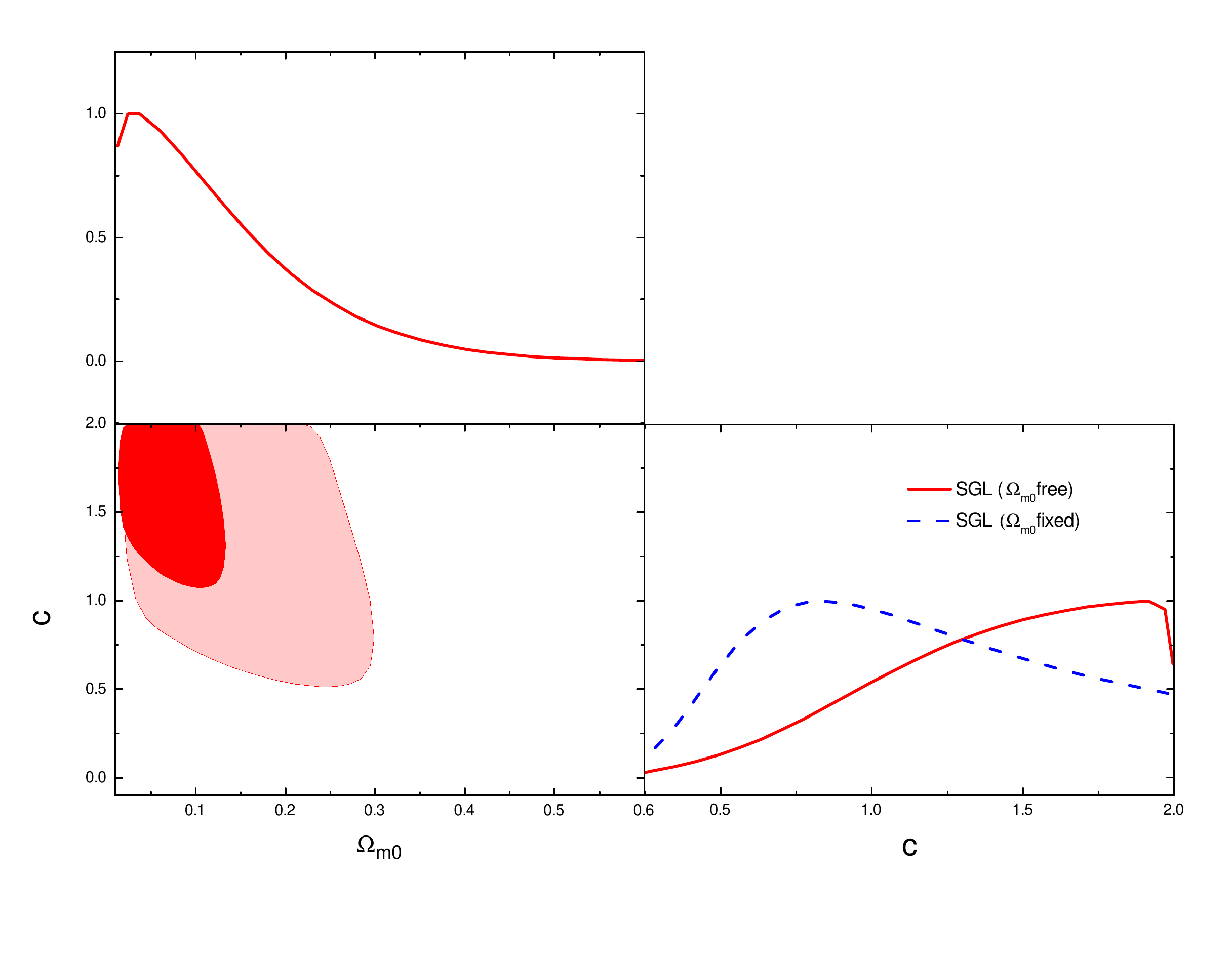}
\caption{\label{fig1} Constraints on the HDE model from the SGL data. In the bottom right panel, 1D marginalized distributions of parameter $c$ with free and fixed $\Omega_{\rm{m0}}$ are presented for comparison.}
\end{figure*}

\begin{figure}[htbp]
\centering
\includegraphics[scale=0.3]{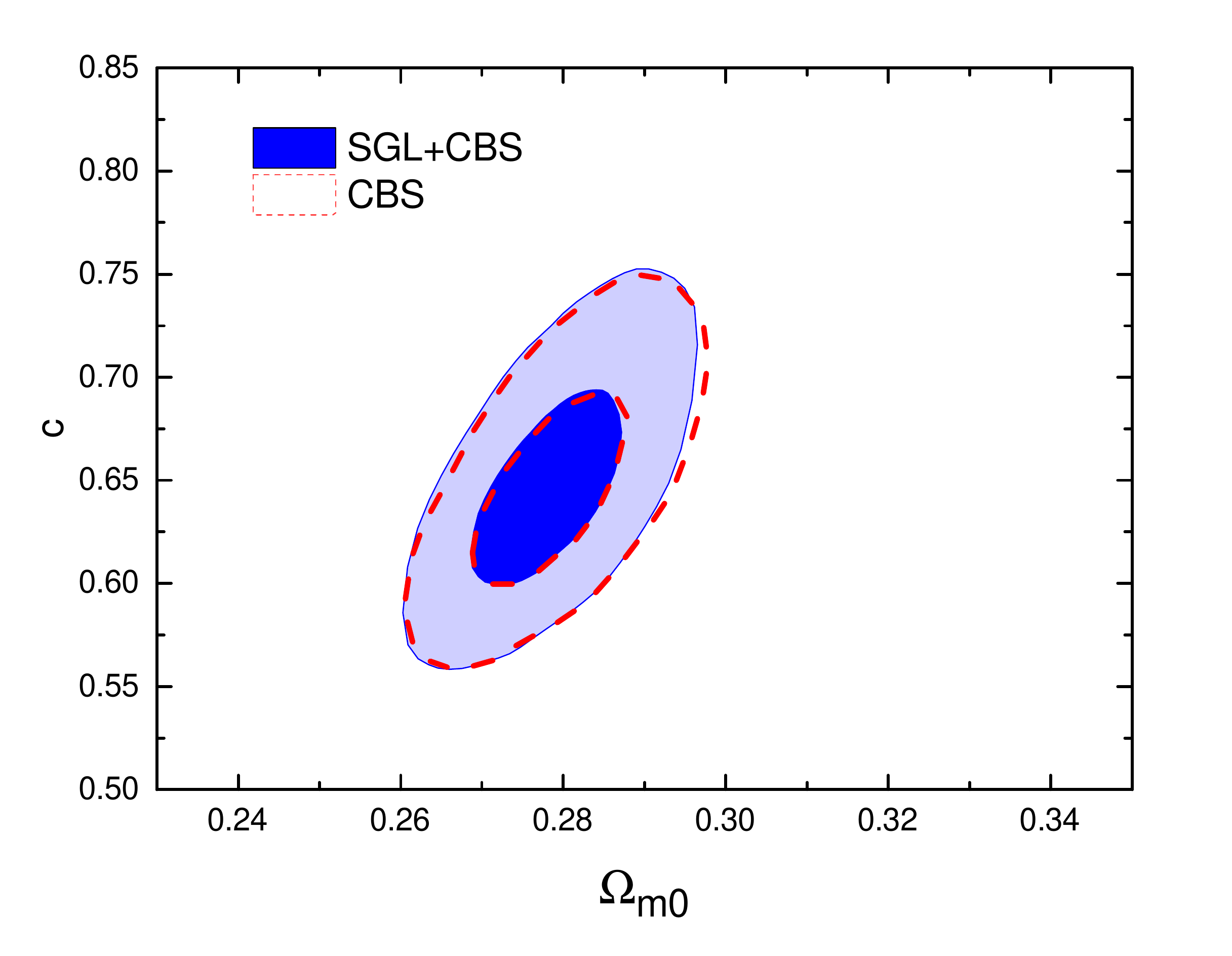}
\caption{\label{fig2} The HDE model: the $1$--$2\sigma$ CL contours in the $\Omega_{\rm{m0}}$-$c$ plane from CBS and the combination of SGL with CBS (CMB+BAO+SN).}
\end{figure}

\section {Constraints on holographic dark energy}\label{sec.3}

In a spatially flat FRW universe consisting of DE (de), matter (m) and radiation (r) (throughout this paper), for HDE with Eqs. (\ref{eq1}) and (\ref{eq2}), the Friedmann equation and the energy conservation equation lead to the following differential equations
\begin{equation}
\frac{1}{E(z)}\frac{dE(z)}{dz}=-\frac{\Omega_{\rm{de}}}{1+z}\left(\frac{1}{2}+\frac{\sqrt{\Omega_{\rm{de}}}}{c}-\frac{\Omega_{\rm{r}}+3}{2\Omega_{\rm{de}}}\right)\label{Ez},
\end{equation}
\begin{equation}
\frac{d\Omega_{\rm{de}}}{dz}=-\frac{2\Omega_{\rm{de}}(1-\Omega_{\rm{de}})}{1+z}\left(\frac{1}{2}+\frac{\sqrt{\Omega_{\rm{de}}}}{c}+\frac{\Omega_{\rm{r}}}{2(1-\Omega_{\rm{de}})}\right),\label{Ode}
\end{equation}
where $\Omega_{\rm{de}}$ is the fractional density of DE, and the fractional density of radiation $\Omega_{\rm{r}}=\Omega_{\rm{r0}}(1+z)^4/E(z)^2$ with $\Omega_{\rm{r0}}=2.469\times10^{-5}h^{-2}(1+0.2271N_{\rm{eff}})$ (the effective number of neutrino species $N_{\rm{eff}}=3.046$). We can obtain $E(z)$ of HDE from numerical solution of differential equations (\ref{Ez}) and (\ref{Ode}) with initial conditions $E_0=1$ and $\Omega_{\rm{de0}}=1-\Omega_{\rm{m0}}-\Omega_{\rm{r0}}$ at $z=0$. When we solve the equations, $\Omega_{\rm{m0}}$ and $c$ are free model parameters.

Firstly, we use $73$ SGL data points to estimate free parameters ($\Omega_{\rm{m0}}$ and $c$) of the HDE model in the light of the public Markov-Chain Monte Carlo package CosmoMC~\cite{Lewis:2002ah}, and get the fit values $\Omega_{\rm{m0}}=0.0244^{+0.1123}_{-0.0136}$ and $c=1.9730^{+0.0270}_{-0.8993}$ (see Table~\ref{table1}). We can see that the best-fit of $\Omega_{\rm{m0}}$ is one order smaller than current observational results. In addition, if $c>1$, the equation of state (EoS) of HDE, $w=-\frac{1}{3}(1+\frac{2}{c}\sqrt{\Omega_{\rm{de}}})$, will be always larger than $-1$ due to $0\leq\Omega_{\rm{de}}\leq 1$, such that the universe avoids entering the de Sitter space-time~\cite{Li:2004rb}. However, a great number of previous constraints favored that $c$ is smaller than $1$, which leads to dark energy behaving as phantom whose EoS $w$ crosses the cosmological constant boundary $-1$ during the evolution. The parameter $c$ is very important to the HDE model.
Furthermore, we plot the $1$--$2\sigma$ confidence level contours in the $\Omega_{\rm{m0}}$-$c$ plane and one-dimentional marginalized probability distributions of $\Omega_{\rm{m0}}$ and $c$ in Fig.~\ref{fig1}, from the SGL data. From Fig.~\ref{fig1}, we can see that it is difficult to effectively constrain parameters $\Omega_{\rm{m0}}$ and $c$, due to the big errors, by using SGL alone. Thus, we consider to calculate the fit value of $c$ in the case of keeping $\Omega_{\rm{m0}}$ at $0.27$, and we get $c=0.8335^{+0.8031}_{-0.3495}$. For direct comparison, we also list the fit results of $c$ in Table~\ref{table1}, and plot 1D marginalized probability distributions of $c$ with free and fixed $\Omega_{\rm{m0}}$ in one plane (the bottom right panel of Fig.~\ref{fig1}). It is found that in the case with fixed $\Omega_{\rm{m0}}$, the value of $c$ is smaller than $1$ , and yet its error is still quite large, which is not a satisfactory result.

Since SGL cannot constrain the parameters of HDE well, we consider to combine it with CBS (CMB+BAO+SN). For comparison, we also estimate the parameters by the CBS only data.  With the CBS data, we get $\Omega_{\rm{m0}}=0.2798^{+0.0083}_{-0.0114}$ and $c=0.6458^{+0.0472}_{-0.0483}$, and with the SGL $+$ CBS data, we get $\Omega_{\rm{m0}}=0.2783^{+0.0092}_{-0.0100}$ and $c=0.6429^{+0.0515}_{-0.0436}$. These results are listed in Table~\ref{table1}. We find that, in the case with SGL $+$ CBS, $c<1$ is at the $6.9\sigma$ level. Moreover, the $1$--$2\sigma$ CL contours in the $\Omega_{\rm{m0}}$-$c$ plane from the CBS and the combination of SGL with CBS data are shown in Fig.~\ref{fig2}. Compared to the CBS data, the SGL+CBS data improve the constraints on $\Omega_{\rm{m0}}$ and $c$ by $3.55\%$ and $0.15\%$, respectively. For the HDE model, the SGL data is helpful to tighten constraints on the parameters.

One possible cause that SGL can improve constraint precision, especially for $\Omega_{\rm{m0}}$, is that it provides some data of high-redshift source. In CBS data, there are $15$ high-redshift data composed of $z_*$ ($z>1000$) from CMB and $14$ SN ($1\leq z\leq1.4$) from SNLS3. Whereas, the SGL data contains $20$ high-redshift sources, in which $5$ sources are in the range of $1\leq z\leq1.4$ and $15$ sources are in the range of $z>1.4$. Owing to supplement high-redshift data, the constraints on model parameters can improve the accuracy.

\section {Comparison with the $w$CDM and RDE models}\label{sec.4}

\begin{table*}\tiny
\caption{Constraint results and corresponding $\chi^2_{\rm{min}}$ for the $w$CDM and RDE models by using the SGL and CBS (CMB+BAO+SN) data, as well as the combination of them. In the case of fixed $\Omega_{\rm{m0}}$, we choose $\Omega_{\rm{m0}}=0.27$.}
\label{table2}
\small
\setlength\tabcolsep{2.8pt}
\renewcommand{\arraystretch}{1.5}
\centering
\begin{tabular}{cccccccccccc}
\\
\hline\hline Model&\multicolumn{3}{c}{$w$CDM} &&\multicolumn{3}{c}{RDE} \\
           \cline{2-4}\cline{6-8}
Parameter   & $\Omega_{\rm{m0}}$ & $w$ & $\chi^2_{\rm{min}}$& & $\Omega_{\rm{m0}}$ & $\alpha$ & $\chi^2_{\rm{min}}$\\ \hline

SGL($\Omega_{\rm{m0}}$ free)
                   & $0.0259^{+0.1331}_{-0.0254}$
                   & $-0.6537^{+0.1663}_{-0.2821}$
                   & $84.1101$&
                   & $0.0159^{+0.0980}_{-0.0058}$
                   & $0.6755^{+0.1221}_{-0.1500}$
                   & $84.1079$                   \\
 SGL($\Omega_{\rm{m0}}$ fixed)
                   & $0.27$
                   & $-0.9772^{+0.2920}_{-0.3812}$
                   & $86.8351$&
                   & $0.27$
                   & $0.4355^{+0.1240}_{-0.1018}$
                   & $88.3526$\\
CBS
                   & $0.2898^{+0.0106}_{-0.0093}$
                   & $-1.0458^{+0.0535}_{-0.0660}$
                   & $425.8848$&
                   & $0.2150^{+0.0071}_{-0.0070}$
                   & $0.3108^{+0.0080}_{-0.0080}$
                   & $599.1870$\\
SGL+CBS
                   & $0.2891^{+0.0100}_{-0.0092}$
                   & $-1.0546^{+0.0606}_{-0.0610}$
                   & $513.0776$&
                   & $0.2153^{+0.0078}_{-0.0058}$
                   & $0.3119^{+0.0082}_{-0.0074}$
                   & $689.5463$\\ \hline
\end{tabular}
\end{table*}

\begin{figure}[htbp]
\centering
\includegraphics[scale=0.3]{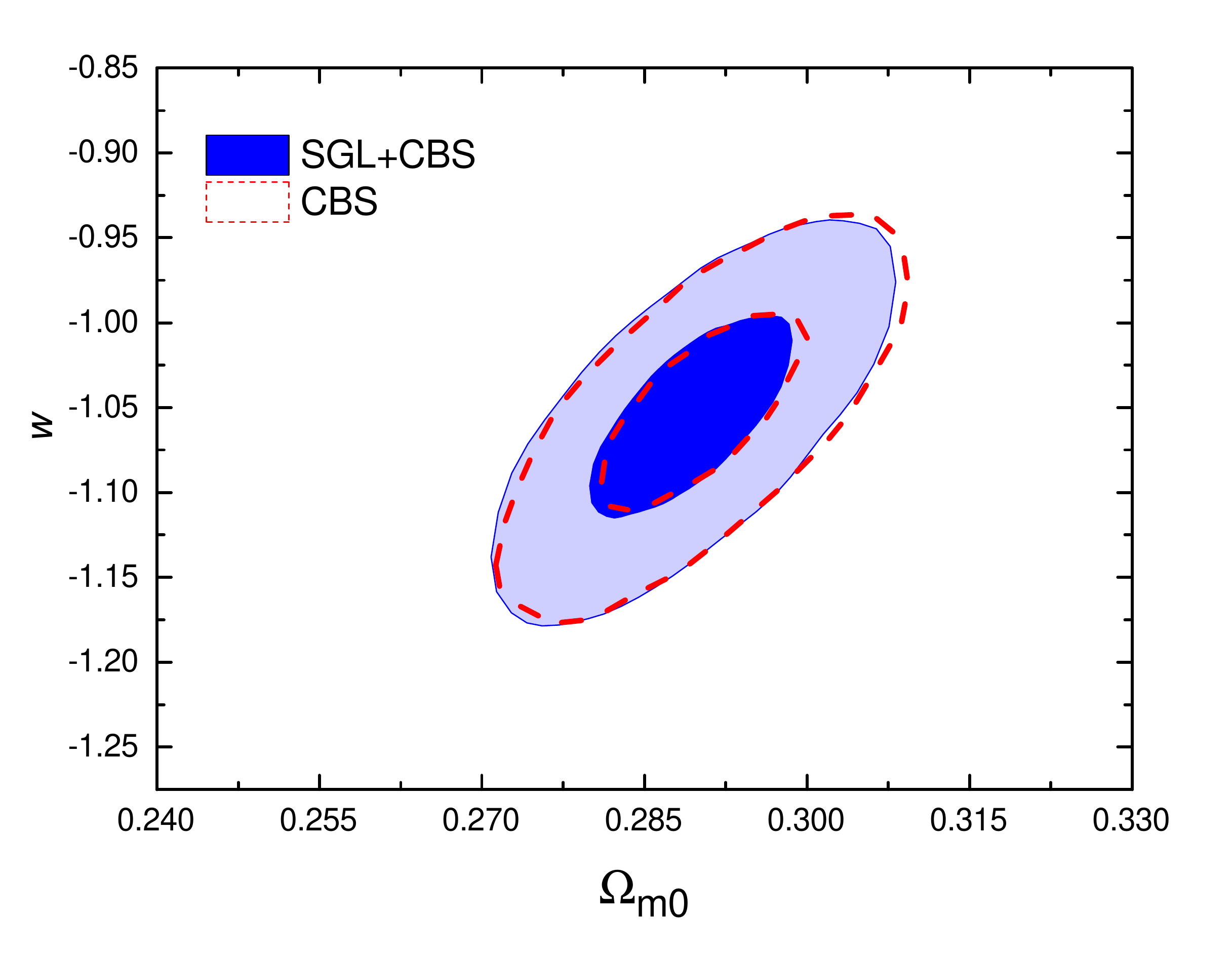}
\caption{\label{fig3} The $w$CDM model: the $1$--$2\sigma$ CL contours in the $\Omega_{\rm{m0}}$-$w$ plane from CBS and the combination of SGL with CBS (CMB+BAO+SN). }
\end{figure}

\begin{figure}[htbp]
\centering
\includegraphics[scale=0.3]{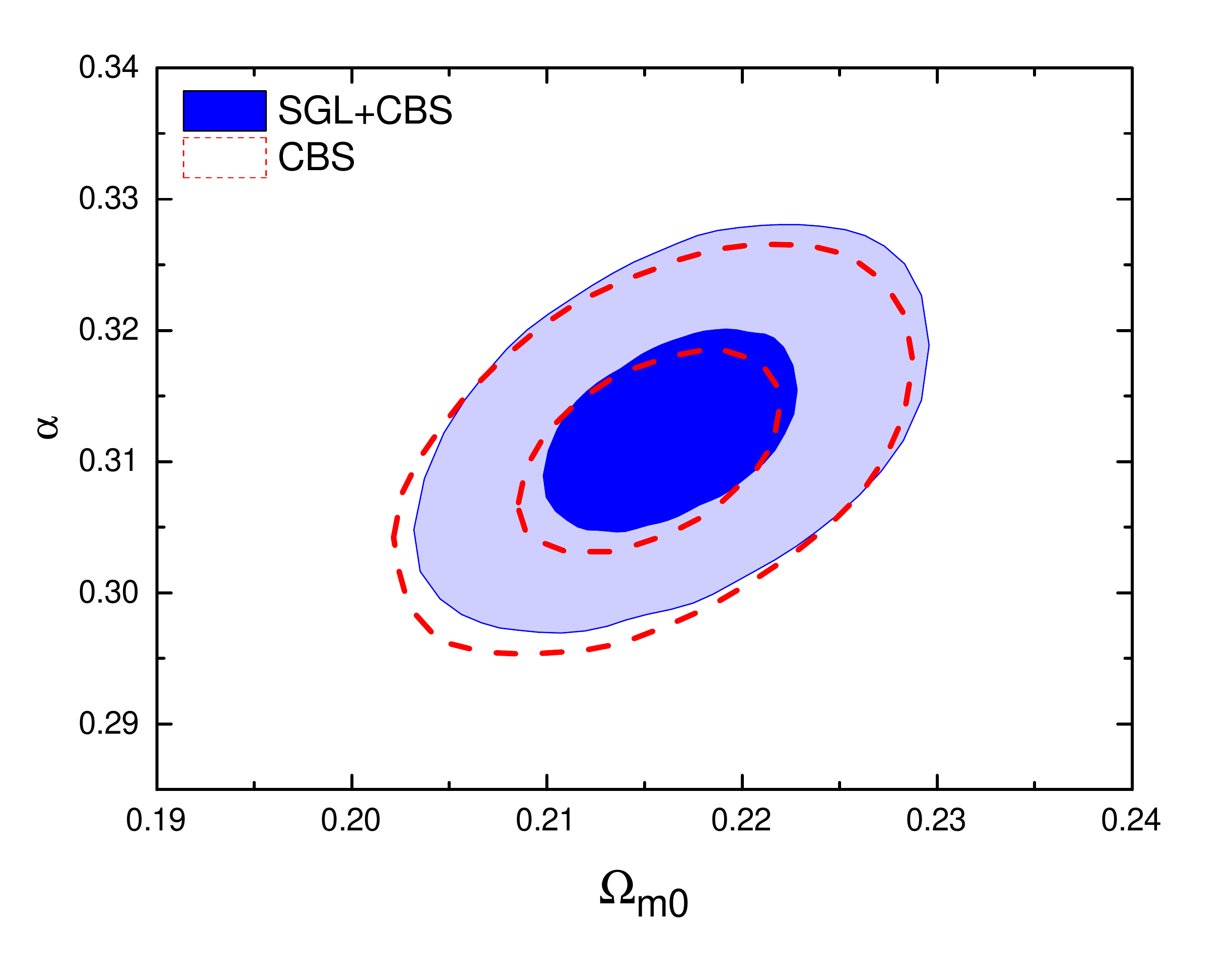}
\caption{\label{fig4} The RDE model: the $1$--$2\sigma$ CL contours in the $\Omega_{\rm{m0}}$-$\alpha$ plane from CBS and the combination of SGL with CBS (CMB+BAO+SN).}
\end{figure}

In this section, we consider the $w$CDM and RDE models, which have the same parameter number with the HDE model.

For the $w$CDM model, $E(z)$ is given by
\begin{equation}\small
E(z)=
\sqrt{\Omega_{\rm{m0}}(1+z)^3+\Omega_{\rm{r0}}(1+z)^4+(1-\Omega_{\rm{m0}}-\Omega_{\rm{r0}})(1+z)^{3(1+w)}}.
\end{equation}

For the RDE model~\cite{Gao:2007ep}, the density of dark energy is also derived from Eq.(\ref{eq1}). But in this case, the IR cutoff $L$ is connected to Ricci scalar curvature, $\mathcal{R}=-6(\stackrel{\centerdot}{H}+2H^2)$.
So Ricci dark energy density is
\begin{equation}
\rho_{\rm{de}}=3\alpha M^2_{\rm{P}}(\stackrel{\centerdot}{H}+2H^2),
\end{equation}
where $\alpha$ is a positive constant. Accordingly, we can get
\begin{equation}\small
E(z)=\sqrt{\frac{2\Omega_{\rm{m0}}}{2-\alpha}(1+z)^3+\Omega_{\rm{r0}}(1+z)^4+(1-\frac{2\Omega_{\rm{m0}}}{2-\alpha}-\Omega_{\rm{r0}})(1+z)^{4-\frac{2}{\alpha}}}.
\end{equation}

Firstly, we use the SGL data to constrain the $w$CDM and RDE models with free and fixed $\Omega_{\rm{m0}}$, and obtain the fit results (see Table~\ref{table2}). From Table~\ref{table2}, we find that SGL only data cannot estimate parameters of the $w$CDM and RDE models well.

Next, we consider to use the CBS data and the combination of SGL with CBS, respectively, to constrain the model parameters and we summarize the corresponding results in Table~\ref{table2}. For the $w$CDM model, the results of the joint constraints are $\Omega_{\rm{m0}}=0.2891^{+0.0100}_{-0.0092}$ and $w=-1.0546^{+0.0606}_{-0.0610}$. Further, the $1$--$2\sigma$ CL contours in the $\Omega_{\rm{m0}}$-$w$ plane for $w$CDM from CBS and the combination of SGL with CBS data are shown in Fig.~\ref{fig3}.
For the RDE model, we get the constraints on $\Omega_{\rm{m0}}$ and $\alpha$ with the SGL $+$ CBS data, $\Omega_{\rm{m0}}=0.2153^{+0.0078}_{-0.0058}$ and $\alpha=0.3119^{+0.0082}_{-0.0074}$, which are improved by $3\%$ and $2.65\%$, respectively, relative to those with CBS only data. In Fig.~\ref{fig4}, we plot the $1$--$2\sigma$ CL contours in the $\Omega_{\rm{m0}}$-$\alpha$ plane for RDE from CBS and the combination of SGL with CBS data.

For the HDE, $w$CDM and RDE models, the values of the total $\chi^2_{\rm{min}}$ are $517.6995$, $513.0776$ and $689.5463$, respectively, with joint SGL $+$ CBS constraints. The HDE model fits these data slightly worse than the $w$CDM model, but much better than the RDE model. Hence, the HDE and $w$CDM models are still competitive models besides the $\Lambda$CDM model.

\section {Conclusion}\label{sec.5}
The strong lensing is an important tool for probing galaxies and cosmology. In this paper, we use $73$ SGL data from SLACS and LSD to constrain the HDE model, as well as the $w$CDM and RDE models.
When only the SGL data are used, for every model, the errors in constraint results are big, and the best-fit of $\Omega_{\rm{m0}}$ is one order smaller than current observations. Even if we fix the value of $\Omega_{\rm{m0}}$ at $0.27$, the errors of another parameter is still big. That is to say, using the SGL data alone cannot constrain the parameters of model well.
However, when the SGL data are combined with CBS (CMB+BAO+SN) data, we get some reasonable results. In the HDE model, we obtain $\Omega_{\rm{m0}}=0.2783^{+0.0092}_{-0.0100}$ and $c=0.6429^{+0.0515}_{-0.0436}$. Particularly, the constraints on $\Omega_{\rm{m0}}$ and $c$ with the SGL $+$ CBS data are improved by $3.55\%$ and $0.15\%$ respectively, compared with those with the CBS data. The same situations also occur in the $w$CDM and RDE models. In the $w$CDM model, with the SGL $+$ CBS constraints, we have $\Omega_{\rm{m0}}=0.2891^{+0.0100}_{-0.0092}$ and $w=-1.0546^{+0.0606}_{-0.0610}$. And, in the RDE model,  with the SGL $+$ CBS constraints, $\Omega_{\rm{m0}}=0.2153^{+0.0078}_{-0.0058}$ and $\alpha=0.3119^{+0.0082}_{-0.0074}$ are improved by $3\%$ and $2.65\%$, respectively, relative to the CBS constraints.
The reason why SGL can tighten constraints in some sense is that it provides some high-redshift data that are important supplementary for the CBS data.
Therefore, we conclude that SGL can be helpful to obtain tighter constraints on model parameters.
Moreover, in terms of the constraints on the HDE model, $c<1$ is at the $6.9\sigma$ level with the SGL $+$ CBS data. According to this result, HDE is likely to become a phantom energy in the future evolution. Through a $\chi^2_{\rm{min}}$ comparison, we conclude that the HDE model is still one of competitive DE models.

\section*{Acknowledgements}

This work was supported by the National Natural Science Foundation of China under Grants No. 11175042 and No. 11522540, the Provincial Department of Education of Liaoning under Grant No. L2012087, and the Fundamental Research Funds for the Central Universities under Grants No. N140505002, No. N140506002, and No. N140504007.

\end{document}